\documentclass{elsart3p}
\usepackage{graphics}
\usepackage{graphicx}
\usepackage{amssymb}
\begin{document}
\begin{frontmatter}
\title{Neutron-scattering study of spin correlations in La$_{1.94-x}$Sr$_{x}$Ce$_{0.06}$CuO$_{4}$}
\author[label1]{M. Enoki}
\ead{enyokin@imr.tohoku.ac.jp }
\corauth{Tel: +81-22-215-2037.  Fax: +81-22-215-2036}
\author[label2]{M. Fujita}
\author[label2]{K. Yamada}
\address[label1]{Department of Physics, Tohoku University, Sendai, Miyagi 980-8578, Japan}
\address[label2]{Institute for Materials Research, Tohoku University, Sendai, Miyagi 980-8577, Japan}
\begin{abstract} 
We performed a neutron-scattering experiment to investigate the effect of distortion of CuO$_2$ planes on the low-energy spin correlation of La$_{1.94-x}$Sr$_{x}$Ce$_{0.06}$CuO$_{4}$ (LSCCO). 
Due to the carrier-compensation effect by co-doping of Sr and Ce, LSCCO has a smaller orthorhombic lattice distortion compared to La$_{2-x}$Sr$_{x}$CuO$_{4}$ (LSCO) with comparable hole concentration $p$.  
A clear gap with the edge-energy of 6$\sim$7 meV was observed in the energy spectrum of local dynamical susceptibility $\chi$$^{\prime\prime}$($\omega$) for both $x$=0.18 ($p$$\sim$0.14) and $x$=0.24 ($p$$\sim$0.20) samples as observed for optimally-doped LSCO ($x$=0.15$\sim$0.18).  
For the $x$=0.14 ($p$$\sim$0.10) sample, in addition to the gap-like structure in $\chi$$^{\prime\prime}$($\omega$) we observed a low-energy component within the gap which develops below 2$\sim$3meV with decreasing the energy.  
The low-energy component possibly coincides with the static magnetic correlation observed in this sample. 
These results are discussed from a view point of relationship between the stability of low-energy spin fluctuations and the distortion of CuO$_2$ planes.
\end{abstract}

\begin{keyword}
High-$T_c$ superconductivity\sep spin fluctuations\sep neutron-scattering  
\end{keyword}
\end{frontmatter}

\section{Introduction} 

Extensive neutron-scattering experiments clarified a close relation between the spin correlations and the superconductivity in high-{\it T}$_c$ superconductors\cite{Kastner}. 
In the prototypical high-{\it T}$_c$ superconductor La$_{2-x}$Sr$_{x}$CuO$_{4}$ (LSCO), in which the doping-dependence of physical properties can be easily investigated by changing Sr concentration, following results were obtained:
(i) the incommensurate (IC) spin correlations were observed in a whole superconducting (SC) phase with 0.055$\leqslant$$x$$\leqslant$0.30\cite{Birgeneau}. 
(ii) the direction of the IC wavevector in the SC phase differs by 45$^{\circ}$ in angle from that observed one in the spin-glass phase for $x$$<$0.055\cite{Matsuda}. 
(iii) the incommensurability ($\delta$) is proportional to {\it T}$_c$ in underdoped region\cite{Yamada}. 
These experimental facts naturally suggest that the origin of IC spin correlations are important to be clarified for the understanding of role of magnetism in the mechanism of high-{\it T}$_c$ superconductivity. 

On the other hand, it is known that the distortion of CuO$_2$ planes can affect the superconductivity and the spin correlations. 
Systematic neutron-scattering of La$_{1.875}$Ba$_{0.125-x}$Sr$_{x}$CuO$_{4}$, in which the crystal structure varies from low-temperature-tetragonal (LTT) phase to low-temperature-orthorhombic (LTO) phase upon increasing $x$, clarified the enhancement of low-energy component of spin fluctuations and the stability of charge stripe order in the LTT phase, where the $T_c$ is well suppressed\cite{Fujita}.
Furthermore, a resistivity measurement performed under high pressures showed the strong correlation between the corrugation of CuO$_2$ planes and $T_c$ in the LTO phase and suggests a higher $T_c$ in the structure with flat CuO$_2$ planes \cite{Nakamura}. 
Therefore, it is interesting to study the relationship between the spin correlation and the corrugation of CuO$_2$ planes in order to extract an intrinsic interplay between the superconductivity and the spin correlation. 

Motivated by the above reason, we have performed neutron-scattering experiment on \\La$_{1.94-x}$Sr$_{x}$Ce$_{0.06}$CuO$_{4}$ (LSCCO), which is co-doped system of Sr$^{2+}$ and Ce$^{4+}$ ions into La$_2$CuO$_4$.
The corrugation of CuO$_2$ planes is relaxed with increasing Sr concentration\cite{Fleming}, and Ce doping is considered to reduce the total hole concentration without a remarkable change in the corrugation of CuO$_2$ planes.
Thereby, we can prepare less-corrugated CuO$_2$ planes in LSCCO compared with LSCO system at similar hole concentration ($p$). 
%
%
%
%
\section{Experimental details}
\begin{figure}[t]
\begin{center}
\includegraphics[width=6.4cm]{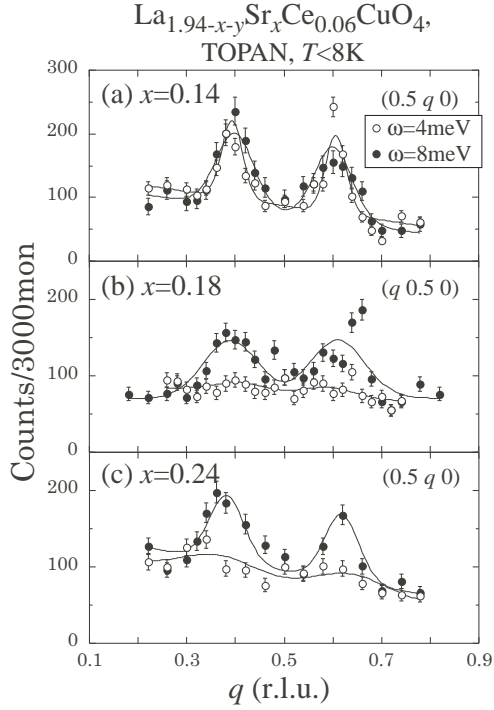}
\end{center}
\caption{The magnetic excitation at $\omega$=4meV and $\omega$=8meV observed in LSCCO (a) $x$=0.14, (b) $x$=0.18, and (c) $x$=0.24 (TOPAN). The vertical scale is normalized by neutron counts monitor. Total counting time corresponds to $\sim$5.9 minutes ($\omega$=4meV) and $\sim$5.4 minutes ($\omega$=8meV). No clear magnetic signal is observed on $x$=0.18 and $x$=0.24, in contrast, well-defined incommensurate peak is observed on $x$=0.14 at $\omega$=4meV
}
\label{f1}
\end{figure}

For the experiment, we prepared the single crystals of LSCCO with $x$=0.14 ($p$$\sim$0.10), $x$=0.18 ($p$$\sim$0.14), and $x$=0.24 ($p$$\sim$0.20), corresponding to underdoped, optimally-doped, and slightly-overdoped sample, respectively. 
The crystals grown by using the traveling-solvent floating-zone method with the typical length of $\sim$100 mm and diameter of $\sim$8 mm were annealed in O$_{2}$ gas flow and cut into $\sim$35mm-length pieces. 
Two of cut samples were assembled so that the CuO$_2$ planes in the horizontal scattering plane, and rest part of samples are further sliced for the magnetic susceptibility measurement. 
%
%
\begin{figure}[t]
\begin{center}
\includegraphics[width=6.4cm]{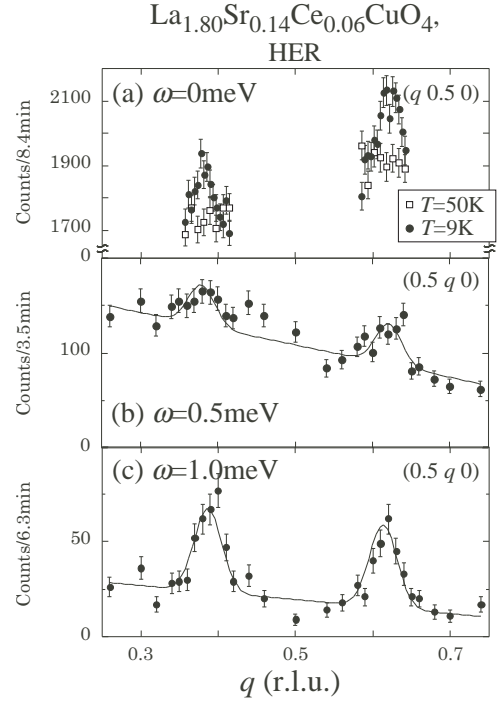}
\end{center}
\caption{
The magnetic spectra (a) elastic, (b) $\omega$=0.5meV, and (c) $\omega$=1.0meV  observed in LSCCO $x$=0.14 (HER). All magnetic spectra show well-defined incommensurate  signals.}
\label{f2}
\end{figure}
%

Inelastic neutron-scattering measurements for 2 meV$\leqslant$$\omega$$\leqslant$12 meV were performed on the thermal neutron triple-axis spectrometer TOPAN installed in the reactor of JRR-3 at Japan Atomic Energy Agency (JAEA). 
We fixed the final neutron energy ($E_f$) at 13.5meV and used a typical horizontal collimation sequence of 50$^{\prime}$-100$^{\prime}$-Sample-60$^{\prime}$-180$^{\prime}$. 
To reduce the flux of high energy neutrons and cut the higher-order wavelengths neutrons, sapphire crystal and the pyrolitic graphite were placed before and after sample, respectively. 
Elastic and low-energy ($\omega$$\leqslant$2 meV) inelastic neutron-scattering measurements with high experimental resolution were done on LSCCO $x$=0.14 by using the cold neutron triple-axis spectrometer HER installed in the guide-hall of  JRR-3. 
In the measurement at HER, $E_f$ of 5meV and the horizontal collimation sequence of 32$^{\prime}$-40$^{\prime}$-S-80$^{\prime}$(180$^{\prime}$)-80$^{\prime}$(180$^{\prime}$) for elastic (inelastic) measurement were selected. 
In the elastic measurement, Be-filter was put before the sample. 
In this paper, crystallographic indexes are denoted as ($h$ $k$ 0) in the tetragonal I4/mmm notation. 

In order to determine the superconducting transition temperature {\it T}$_c$, we measured the magnetic susceptibility with using a superconducting quantum interference device magnetometer. 
Evaluated {\it T}$_c$'s were $\sim$30K ($x$=0.14), $\sim$35K ($x$=0.18), and $\sim$25K ($x$=0.24), respectively. 
We, furthermore, determined $T_{\rm c}$'s for several crystals with the $x$-range from 0.10 to 0.30 to make a phase diagram for the LSCCO system.
From a comparison of phase diagram of the present LSCCO and LSCO systems, we concluded that 6\% Ce doping reduces $\sim$4\% hole concentration. 
In addition, it was confirmed by our neutron-scattering measurement that the Ce-doping does not affect both degree of in-plane orthorhombic distortion at low temperature and the structural transition temperature from high-temperature-tetragonal (HTT) to LTO $T_{\rm d1}$. 
Therefore, the superconductivity occurs on the less-corrugated CuO$_2$ planes in LSCCO compared with LSCO with similar hole concentration. 
Detailed results will be presented elsewhere\cite{enk}.


\section{Results and Discussion}
Figure 1 shows the constant-$\omega$ spectra measured at TOPAN.
Well-defined IC peaks were observed in all samples at $\omega$=8 meV (closed circles). 
In contrast, no clear magnetic signal was observed in the spectra measured at $\omega$=4 meV (open circles) for the $x$=0.18 and $x$=0.24,
while the $x$=0.14 sample shows sharp IC peaks at this energy. 
The incommensurability ($\delta$), corresponding to the half distance between the two peaks, is $\sim$ 0.12 r.l.u. in all samples. (1 r.l.u. corresponds to 1.67 ${\rm \AA}$$^{-1}$. ) 
Thus, $\delta$ is independent on $x$ in the measured concentration range.  
Absence of well-defined magnetic signal at low-energy in the $x$=0.18 and 0.24 samples is consistent with the opening of a spin-gap, which is reported for the optimally-doped LSCO. 
On the other hand, the persistence of magnetic signal with $\omega$ down to 2 meV in the $x$=0.14 sample is rather analogous to the results for the underdoped LSCO exhibiting a short range magnetic order at low temperature. 

We next pay our attention to the low-energy spin fluctuations below 2 meV and the possible static magnetic order. 
Elastic magnetic signal was searched for all samples using the thermal neutron spectrometer and the signal was observed only in the $x$=0.14 sample. 
Then we investigated the low-energy and the static spin correlations in the $x$=0.14 with higher experimental resolution ($\Delta$$\omega$$\sim$0.2meV and $\Delta$$Q$$\sim$0.005${\rm \AA}^{-1}$) at the cold triple-axis spectrometer. 
Figure 2 shows the elastic and the representative inelastic spectra. 
Existence of IC magnetic peaks at $\omega$=0.5meV suggests a gap-less spin excitation spectrum in the $x$=0.14 sample (Fig.2 (b)). 
Furthermore, as seen in Fig. 2(a), the elastic magnetic intensity was reconfirmed to appear at low temperatures. 
From the temperature-dependence of peak-intensity, the onset temperature for the appearance of intensity ($T_m$) was determined to be $\sim$20 K. 
%

\begin{figure}[t]
\begin{center}
\includegraphics[width=7cm]{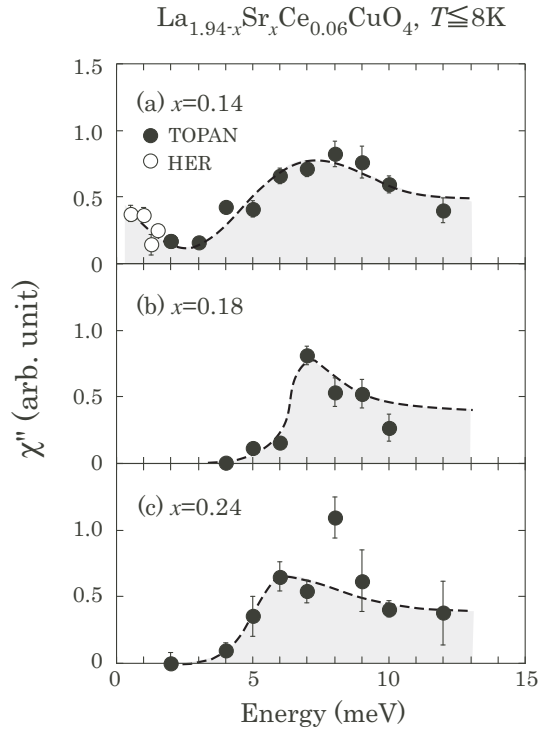}
\end{center}
\caption{ The excitation energy-dependence of $\chi$$^{\prime\prime}$($\omega$) on (a)$x$=0.14, (b)$x$=0.18, and (c)$x$=0.24. Filled circle and open circle represents using the data on TOPAN and HER, respectively.  Clear gap structure exist in spin excitation on $x$=0.18 and $x$=0.24. Otherwise, $x$=0.14 has gap-like structure in the magnetic excitation.}
\label{f4}
\end{figure}

For the evaluation of local spin susceptibility  $\chi$$^{\prime\prime}$($\omega$), which corresponds to $Q$-integrated intensity after correcting the thermal factor, we take into account four isotropic peaks at (0.5$\pm$$\delta$ 0.5 0) and (0.5 0.5$\pm$$\delta$ 0) in the analysis. 
Then the following Gaussian function was fitted to the observed spectra with convoluting experimental resolution, 
%
\begin{displaymath}
\frac{{\chi}^{\prime\prime}(\omega)} {1-exp(-\hbar\omega/{\it k}_{B}{\it T})}
exp(-ln(2)(\mbox{\boldmath $Q$}_{AF}\pm\mbox{\boldmath $q$}_{\delta})^2/\kappa^2),
\end{displaymath}
where {\boldmath $Q$}$_{AF}$ and {\boldmath $q$}$_{\delta}$ represent AF wavevectors of (0.5 0.5 0) and IC wavevectors of ($\delta$ 0 0)/(0 $\delta$ 0), respectively, and  
$\kappa$ is the peak-width (half-width at half-maximum). 
$\omega$-dependence of $\chi$$^{\prime\prime}$($\omega$) for three samples is plotted in Fig. 3. 
The results obtained for the $x$=0.14 sample at thermal and cold spectrometers are normalized with the values at 2 meV. 
The magnetic excitation spectra in the $x$=0.18 and $x$=0.24 samples show a clear gap-structure, while $\chi$$^{\prime\prime}$($\omega$) in the $x$=0.14 sample has finite values in the all measured $\omega$-range below 12 meV.
In the latter, though the decrease of $\chi$$^{\prime\prime}$($\omega$) below $\sim$ 7 meV is resemble to the gap-structure observed in the $x$=0.18 and 0.24 samples, the intensity start to develop below 2 meV with decreasing $\omega$. 
%
%

Here, we discuss the structural effects on the spin correlations from the comparison between the results for LSCCO and LSCO with comparable hole concentrations. 
First, we consider the gap-structure in the LSCCO with $x$=0.18 and 0.24 samples. 
The energy spectra for both samples are nearly the same with that for the optimally-doped LSCO, although the in-plane lattice distortion is smaller in the LSCCO system\cite{Lee03}. 
%
%
Therefore, the gap-structure near the optimally doped region is not affected by a small lattice distortion.

%
We note that the appearance of the energy-gap by Ce doping into the Sr- overdoped sample is one of strong evidence of the reduction of hole concentration by the formation of Ce$^{4+}$ state. We hence emphasize that the absence of spin-gap in the overdoped LSCO ($x$$\sim$0.25)\cite{Lee00,Wakimoto} is not an extrinsic chemical disorder effect of dopants but an inherent property in overdoped region.

Second, the gap-like structure observed in the underdoped LSCCO with $x$=0.14 is contrastive to the gapless structure in the LSCO with $x$=0.10, although the hole concentration in these two samples is comparable. 
Difference in the $\chi$$^{\prime\prime}$($\omega$) between the two samples is prominent in the low-energy part below around 7 meV. 
%
%
%
Therefore the absence of spin-gap in the underdoped LSCO sample can be ascribed by the slowing down of spin fluctuations.
We speculate carrier localization on the corrugated CuO$_2$ planes is one of the main reasons for the slowing down of spin fluctuations. The stronger carrier localization occurs when the carrier concentration is smaller.
When spin and charge stripes are formed, the carrier localization corresponds to the pinning of stripes by the lattice potential on the corrugated CuO$_2$ planes. 
Pinning of stripes could take place even in the LTO phase, although the pinning effect in the LTO phase is weaker than that in the LTT phase. 
%
%
In this context, the observed gap-like structure in $\chi$$^{\prime\prime}$($\omega$) for the LSCCO with $x$=0.14 sample suggests that spin-gap states intrinsically exist even in the underdoped La214 system, which  easily becomes unclear due to the structural effect. 
%
%
%
The low-energy spin fluctuations below 2 meV and the static short range order observed in the present LSCCO with the $x$=0.14 sample would be induced by residual corrugation of CuO$_{2}$ planes and/or disorder of large amount of doped Sr and Ce ions\cite{Goko}.
Evidence for the existence of the low-eneryg spin fluctuations below the gap energy in the LSCO system is recently reported from the inelastic neutron scattering measurement done under magnetic fields\cite{Chang}.
To clarify the intrinsic spin correlations in the high-$T_c$ superconductors, further systematic neutron-scattering experiments are needed.

\end{document}